# Excitonic Stark Response Induced by Quantum Capacitance in van der Waals Heterostructures

Xiangzhi Hong[1], Yongkang Jian[1], Yang Chen[1], Rukai Cang[1], Yuanlong Guo[1], Takashi Taniguchi[2], Kenji Watanabe[3], Baigeng Wang[1,4,5], and Erfu Liu[1,4,5*]

[1]National Laboratory of Solid State Microstructures, School of Physics, Nanjing University, Nanjing 210093, China

[2]Research Center for Materials Nanoarchitectonics, National Institute for Materials Science, 1-1 Namiki, Tsukuba 305-0044, Japan

[3]Research Center for Electronic and Optical Materials, National Institute for Materials Science, 1-1 Namiki, Tsukuba 305-0044, Japan

[4]Collaborative Innovation Center of Advanced Microstructures, Nanjing University, Nanjing 210093, China

[5]Jiangsu Physical Science Research Center, Nanjing 210093, China

[*]Contact author: erfuliu@nju.edu.cn

**Abstract:**

Precise electrostatic control in van der Waals heterostructures is commonly described by a geometric capacitor model, which assumes that the internal electric field is fully determined by external gate voltages. Here, we show that this description breaks down in the presence of interfacial charge transfer, where quantum capacitance modifies the internal electrostatics. Using interlayer excitons as local probes, we observe a dipole-dependent energy splitting that persists even when the gate-induced geometric field is nominally canceled. This splitting reflects incomplete cancellation between gate- and carrier-induced fields set by quantum capacitance, while many-body renormalization is largely symmetric with respect to the two opposite dipole orientations, allowing the splitting to provide a direct optical measure of the local chemical potential. Our results reveal the central role of quantum capacitance in interface electrostatics and provide a route to probe it in operando.

Electrostatic gating is central to modern two-dimensional (2D) electronics and optoelectronics, enabling precise control of carrier density and out-of-plane electric fields without introducing chemical impurities. In particular, the dual-gate architecture has become a standard platform for investigating many-body physics in van der Waals (vdW) heterostructures. By independently tuning the top and bottom gate voltages, this geometry allows decoupled control of the carrier density $n$ and the out-of-plane electric field $E$. This capability is essential for studying a variety of field- and doping-dependent phenomena, including the Stark effect of interlayer excitons (IXs) in transition metal dichalcogenides [1–15], topological phase transitions [16–21], and tunable correlations and magnetism in moiré superlattices [22–36]. In most cases, the internal electrostatic environment in these devices is described by a geometric capacitor model, where the electric field is determined solely by the applied voltages and the dielectric properties of the layered structure [37–40].

However, carrier doping inevitably introduces additional electrostatic fields and screening effects. When the carrier distribution remains symmetric across the structure, the carrier-induced internal electric fields largely cancel, preserving the apparent validity of the geometric model [Figs. 1(a) and 1(b)]. When this symmetry is broken by interfacial charge transfer, however, the geometric description breaks down in two ways [Figs. 1(c) and 1(d)]. First, the redistributed charges generate additional internal fields that modify the total electrostatic profile. Second, the finite density of states of the interfacial layer gives rise to quantum capacitance, which alters both the local chemical potential and the effective field distribution. As a result, the internal electrostatics can deviate substantially from the prediction of the geometric capacitor model. Direct experimental access to this electrostatic reconstruction remains challenging.

In this *Letter*, we use interlayer excitons in bilayer $MoS_2$ as sensitive, *in situ* probes to quantify the complex electrostatic environment. Unlike global transport measurements that require sophisticated device fabrication, this optical approach enables straightforward, micrometer-scale probing of local properties. A key challenge is that the exciton energy shift contains contributions from both the Stark effect and many-body screening arising from doped carriers, making it challenging to identify the underlying mechanism. We address this by exploiting the distinct symmetry of the two contributions with respect to the two opposite out-of-plane dipole orientations. First, in the top-gate-scan configuration, the gate-induced and carrier-induced fields oppose each other, canceling the net field variation predicted by the geometric model. Second, a symmetry-based analysis isolates the residual Stark response from dipole-symmetric many-body renormalization. Specifically, we show that many-body screening produces nearly identical energy shifts for the two opposite-dipole configurations, whereas the residual Stark response appears as a dipole-dependent splitting. This dual-level strategy enables direct extraction of the intrinsic interfacial electrostatics beyond the conventional geometric capacitor model.

Our devices are fabricated by stacking monolayer graphene on bilayer $MoS_2$ in a dual-gate architecture. The heterostructures are encapsulated by hexagonal boron nitride (hBN) flakes of comparable thickness, while few-layer graphite flakes serve as the contact and gate electrodes [Fig. 2(a)]. Fabrication details are provided in Note 1 of Supplemental Material [41] (see also Refs. [42–47] therein). Because of the band alignment between graphene and bilayer $MoS_2$, charge carriers transfer to the graphene layer. This charge redistribution generates an additional internal electric field and modifies excitonic interactions through many-body screening. As a result, the energies of the two interlayer exciton (IX) branches with opposite dipole moments shift in bilayer $MoS_2$. As shown in Fig. 2(b), these IX states acquire oscillator strength through hybridization of the valence bands [9–12]. Based on this device structure, we probe the effects of transferred carriers through reflection-contrast spectroscopy of the interlayer excitons (see Note 2 in Supplemental Material [41] for measurement details). All measurements were performed at approximately 6 K.

Figures 2(c) and 2(d) show reflection-contrast maps of bilayer $MoS_2$ as functions of out-of-plane electric field at charge neutrality and carrier density at zero applied gate field, respectively (see Note 3 of Supplemental Material [41] for details on the calculations of carrier density and electric field). In Fig. 2(c), the electric-field-dependent map resolves the 1s interlayer exciton states with opposite dipole orientations, denoted $\mathrm{IX}_{\downarrow}^{1s}$ and $\mathrm{IX}_{\uparrow}^{1s}$, where the subscript arrows indicate the direction of the out-of-plane electric dipole moment, as well as the 1s intralayer exciton, denoted $\mathrm{X}^{1s}$. From Fig. 2(c), we extract the energies of the two 1s interlayer exciton branches and plot them in Fig. 2(e). The energies vary linearly with the electric field, consistent with a linear Stark shift arising from their opposite out-of-plane dipole moments. By fitting the energy dispersion as a function of electric field, we obtain a dipole moment of approximately $0.48\ e\,\mathrm{nm}$ for the two interlayer exciton branches, in good agreement with previous reports [9–11]. This result confirms that the observed energy shifts are dominated by the electric-field-induced Stark effect for the interlayer excitons.

In Fig. 2(f), we plot the interlayer exciton energies extracted from Fig. 2(d) as a function of carrier density. The exciton energies shift with increasing carrier density, even though the net external electric field imposed by the dual gates is maintained at zero [Fig. 1(d)]. This behavior indicates that charge transfer induces a non-negligible internal electric field acting on the interlayer excitons. To quantify this effect, we calculate the expected Stark shift by incorporating the carrier-induced internal field ($E_{\mathrm{carrier}}$) into the standard geometric capacitor model (see Note 4 of Supplemental Material [41]). Although the calculation reproduces the qualitative trend of the experimental data, a clear quantitative discrepancy remains, as shown in Fig. 2(f). This deviation indicates that the electrostatics cannot be fully captured by the geometric model alone, pointing to additional contributions such as quantum capacitance and many-body screening.

To probe the limitations of the geometric model and isolate nongeometric contributions, we perform a control experiment based on single-gate voltage scans. In this configuration, the relative alignment between the gate-induced external field $E_{\text{gate}}$ and the carrier-induced internal field $E_{\text{carrier}}$ is reversed: they are parallel during the bottom-gate scan and antiparallel during the top-gate scan [Figs. 3(a) and 3(b)]. This distinction allows us to selectively enhance or suppress the net electric field experienced by the interlayer excitons ($E_{\text{net}}$). Specifically, we sweep the bottom-gate voltage while keeping the top gate fixed at 0 V, and sweep the top-gate voltage while fixing the bottom gate at $-0.6\ \text{V}$. The resulting reflection-contrast maps and the extracted exciton energies are shown in Figs. 3(c)–3(f).

During the bottom-gate sweep, $E_{\text{gate}}$ and $E_{\text{carrier}}$ are parallel and add constructively. Although deviations from the geometric model are present, they are masked by the large Stark shift associated with the geometric field, resulting in a monotonic evolution similar to that in Fig. 2(f). In contrast, during the top-gate sweep, $E_{\text{gate}}$ and $E_{\text{carrier}}$ are antiparallel. Within the geometric capacitor model, their variations are equal in magnitude and opposite in direction, leading to cancellation of the net geometric field variation across the sweep (see Note 5 in Supplemental Material [41]). The geometric Stark background is therefore removed within this model.

In this regime, any remaining spectral response reflects contributions beyond the geometric model. As shown in Fig. 3(f), the extracted exciton energies exhibit a pronounced asymmetry with respect to both dipole orientation and carrier polarity: $\text{IX}_{\downarrow}^{1\text{s}}$ shows a weak shift below the Dirac point but a strong redshift above it, while $\text{IX}_{\uparrow}^{1\text{s}}$ displays the opposite trend. This behavior depends on the relative carrier types in graphene and the $MoS_2$ layer adjacent to it: a pronounced redshift occurs when the carrier type in graphene matches the dominant carrier type in the adjacent $MoS_2$ layer, whereas only a weak shift is observed when they differ.

The exciton energy shifts therefore depend on both dipole orientation and carrier polarity. In general, the spectral response of interlayer excitons contains two contributions: the electrostatic Stark effect determined by the net electric field, and many-body screening induced by doped carriers [Fig. 3(g)]. The overall redshift suggests the presence of a many-body contribution; however, its role in the dipole-dependent asymmetry remains unclear.

Two possible mechanisms can account for this asymmetry. It may originate from a residual Stark effect due to incomplete cancellation of the opposing electric fields, or from a dipole-dependent many-body contribution. In the latter case, many-body screening affects the exciton energy through both bandgap renormalization (BGR) and a reduction of the exciton binding energy [48–53]. Distinguishing between these scenarios therefore requires determining whether these many-body effects contribute differently to the two opposite-dipole branches. In the following, we analyze the exciton fine structure to separate these contributions and identify the dominant mechanism.

To identify the origin of the nongeometric deviation observed in the antiparallel configuration, we focus on the top-gate-dependent data. Since the oscillator strength of interlayer excitons arises from valence band hybridization in bilayer $MoS_2$, their layer-asymmetric electron-hole wavefunction distribution could, in principle, lead to different many-body renormalizations for the two opposite dipole orientations [9,10,12].

We first examine whether the observed asymmetry originates from changes in the exciton binding energy $E_b$. Figure 4(a) shows the second-derivative reflection-contrast map of the 2s exciton states, where four distinct peaks corresponding to $X_1^{2s}$, $X_2^{2s}$, $IX_{\downarrow}^{2s}$, and $IX_{\uparrow}^{2s}$ are clearly resolved, with the arrows indicating the out-of-plane dipole orientations. We extract the energies of the 1s and 2s states for both interlayer exciton branches and plot them in Fig. 4(b). The 2s interlayer excitons exhibit an asymmetry similar to that of the 1s states with respect to both dipole orientation and carrier polarity.

To quantify the evolution of the binding energy, we evaluate the change in the energy separation between the 2s and 1s states relative to the charge-neutral point, $\Delta E_{2s-1s}$, which serves as a proxy for the binding energy variation under the assumption of approximate Rydberg scaling [54–61]. As shown in Fig. 4(c), $\Delta E_{2s-1s}$ decreases by approximately 5–7 meV over the measured doping range and exhibits nearly identical behavior for the two interlayer exciton branches with opposite dipole orientations. This trend and energy scale are comparable to previous observations of graphene-induced dielectric screening in $WS_2$-graphene heterostructures [62], although the material and device configurations are different. The nearly identical responses of the two opposite-dipole branches indicate that the binding-energy renormalization is largely symmetric with respect to dipole orientation and therefore cannot account for the dipole-dependent spectral asymmetry.

Next, we consider bandgap renormalization (BGR), which contains Coulomb-hole and screened-exchange contributions [63–65]. The Coulomb hole arises from long-range dielectric screening and is expected to affect the two interlayer exciton branches similarly. The screened-exchange contribution should be relatively weak in our geometry because the doping carriers reside primarily in graphene and have little wavefunction overlap with $MoS_2$. Taken together, these considerations indicate that BGR is expected to be largely symmetric with respect to dipole orientation and therefore unlikely to account for the observed dipole-dependent asymmetric spectral response.

While we cannot exclude a small dipole-dependent many-body contribution below our experimental resolution, the near-identical evolution of $\Delta E_{2s-1s}$ for the two branches, together with the suppressed screened-exchange contribution expected when the doping carriers reside primarily in graphene, indicates that the observed asymmetry is dominated by the residual Stark response arising from incomplete cancellation of the opposing electric fields. This incomplete cancellation originates from the finite density of states of graphene and the associated quantum capacitance. For a given applied gate voltage, the quantum capacitance acts in series with geometric capacitance and reduces the transferred carrier density relative to the prediction of the geometric capacitor model. Consequently, the carrier-induced internal field does not fully compensate for the external gate field, leaving a residual electric field that produces the observed Stark shift. This behavior provides direct evidence of quantum-capacitance-limited electrostatics in the present vdW heterostructure.

To quantitatively test this picture, we track the change in the energy splitting between the two 1s interlayer exciton branches with opposite dipole orientations relative to charge neutrality. This differential quantity suppresses contributions that are largely symmetric with respect to dipole orientation, including binding-energy and bandgap renormalization. We then calculate the residual Stark splitting using an electrostatic model that includes the quantum capacitance of graphene $C_q$ (see Note 6 in Supplemental Material [41]). The calculated dependence on top-gate voltage agrees well with the experimental splitting in Fig. 4(d), without introducing additional fitting parameters.

This agreement establishes a direct relation between the optical response and the local chemical potential. In the antiparallel configuration with the bottom-gate voltage fixed, the change in the residual electric field is determined by the change in the potential of graphene relative to the fixed bottom gate. Consequently, the measured energy splitting scales linearly with the graphene chemical potential (see Note 6 in Supplemental Material [41]). This linear dependence allows us to establish an experimental calibration (see Note 7 in Supplemental Material [41])—without relying on multiparameter fitting— to convert the optical Stark splitting into a quantitative measure of the local chemical potential [Fig. 4(d)].

The consistency of this interpretation is supported by our two-step analysis. The antiparallel gate configuration removes the geometric Stark background expected from the geometric capacitor model, while the energy difference between the two opposite-dipole branches suppresses many-body contributions that are largely symmetric with respect to dipole orientation. Similar behavior is observed at different fixed bottom-gate voltages and in device 2 (see Notes 8 and 9 in Supplemental Material [41]). In device 3, the bilayer-graphene region agrees well with the corresponding calculation, whereas the multilayer-graphene region shows only a weak top-gate dependence within our experimental resolution (see Note 10 in Supplemental Material [41]). These results show that the residual response depends on the density of states of the charge-receiving layer. More generally, quantum-capacitance corrections become important when interfacial charge transfer occurs, and the quantum capacitance is comparable to the relevant geometric capacitance. Under these conditions, the chemical-potential shift modifies the relation between gate voltage and transferred charge, leading to deviations from the geometric capacitor model.

In summary, we use interlayer excitons as sensitive *in situ* probes to investigate the electrostatics of graphene-semiconductor interfaces. We show that the observed spectral asymmetry arises from a residual internal electric field governed by the quantum capacitance of graphene. The dipole-dependent component provides a measure of the local chemical potential, while the dipole-symmetric component tracks many-body screening at the interface (see Fig. S5 [41]). Unlike previous approaches that rely on feedback loops or dielectric isolation [66], our method provides a direct optical readout and remains robust in the presence of complex dielectric environments. These results highlight the role of quantum capacitance in determining internal electric fields in vdW heterostructures and establish excitonic spectroscopy as a versatile probe of buried interfaces under operating conditions.

**Acknowledgments—**E. L. acknowledges support from the National Key R&D Program

of China (Grant No. 2024YFA1410500), the National Natural Science Foundation of China (Grant No. 12374456), the Program for Innovative Talents and Entrepreneur in Jiangsu (Grant No. JSSCTD202408), the Jiangsu Provincial Department of Science and Technology (Grant No. BK20253040), and the Fundamental Research Funds for the Central Universities. B. W. acknowledges support from the National R&D Program of China (Grant No. 2022YFA1403601), the Natural Science Foundation of Jiangsu Province (Grant No. BK20233001). K. W. and T. T. acknowledge support from the JSPS KAKENHI (Grants No. 21H05233 and No. 23H02052), the CREST (JPMJCR24A5), JST, and World Premier International Research Center Initiative (WPI), MEXT, Japan.

X. H. and R. C. fabricated the devices; X. H., Y. J., Y. C., and Y. G. performed the measurements. X. H. and E. L. analyzed the data. B. W. contributed to the theoretical explanation. T. T. and K. W. provided boron nitride crystals for device fabrication. E. L. supervised the research and coordinated the work. E. L. and X. H. wrote the manuscript.

The authors declare no competing interests.

**Data availability—**The data that support the findings of this article are not publicly available. The data are available from the authors upon reasonable request.

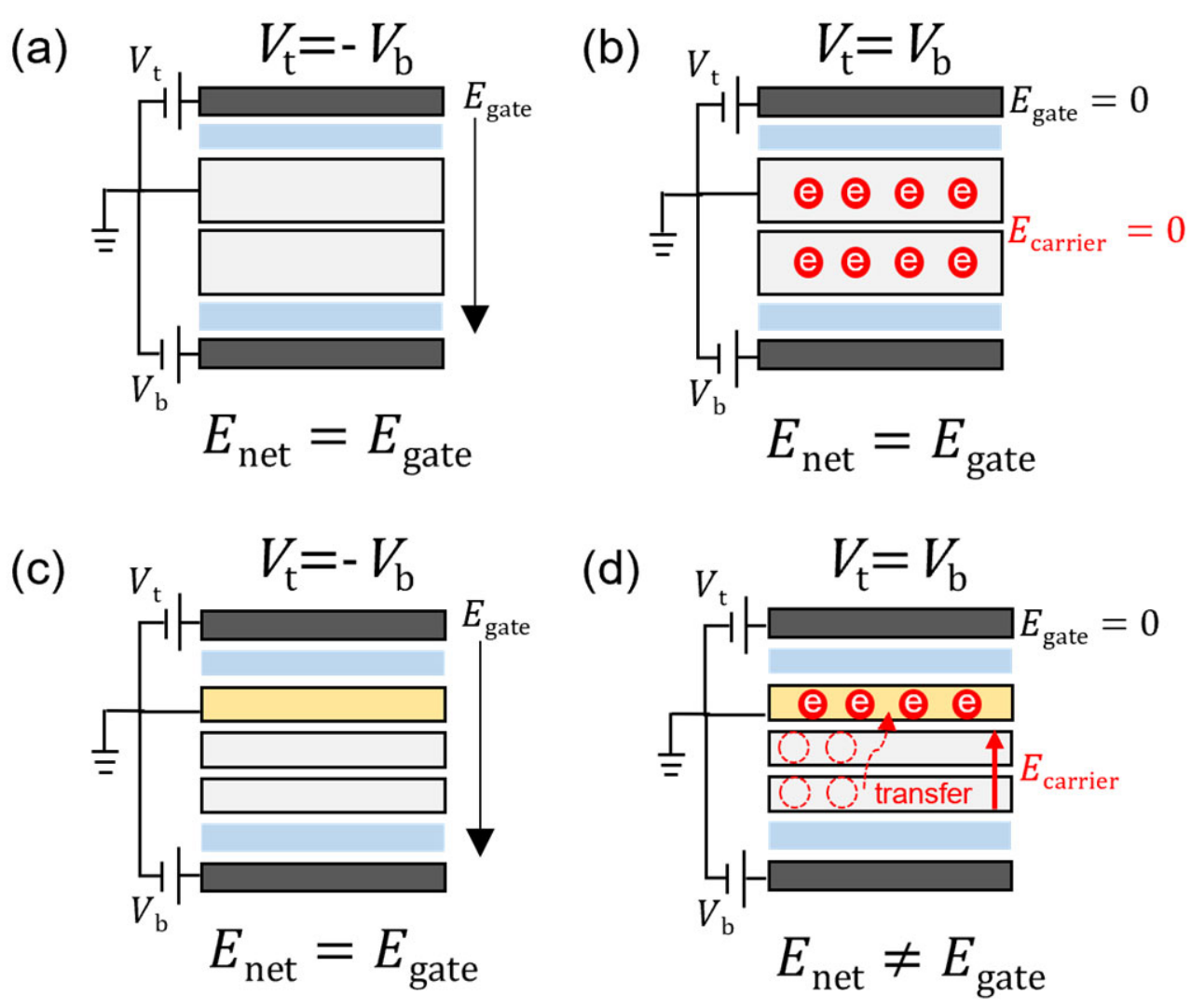

**FIG. 1.** Breakdown of the geometric capacitor model induced by charge transfer. Schematic illustrations of symmetric device geometries in (a) and (b) and asymmetric device geometries in (c) and (d). In (a)-(c), the net electric field $E_{\mathrm{net}}$ equals the applied gate field $E_{\mathrm{gate}}$, indicating that the geometric capacitor model remains applicable. This occurs in the undoped cases (a) and (c) because no free carriers are present, and in the symmetrically doped case (b) because the carrier-induced fields $E_{\mathrm{carrier}}$ cancel owing to their symmetric distribution. In contrast, asymmetric doping caused by interfacial charge transfer (d) produces an imbalanced carrier distribution and hence a finite $E_{\mathrm{carrier}}$ even at $E_{\mathrm{gate}} = 0$, leading to the breakdown of the geometric capacitor model.

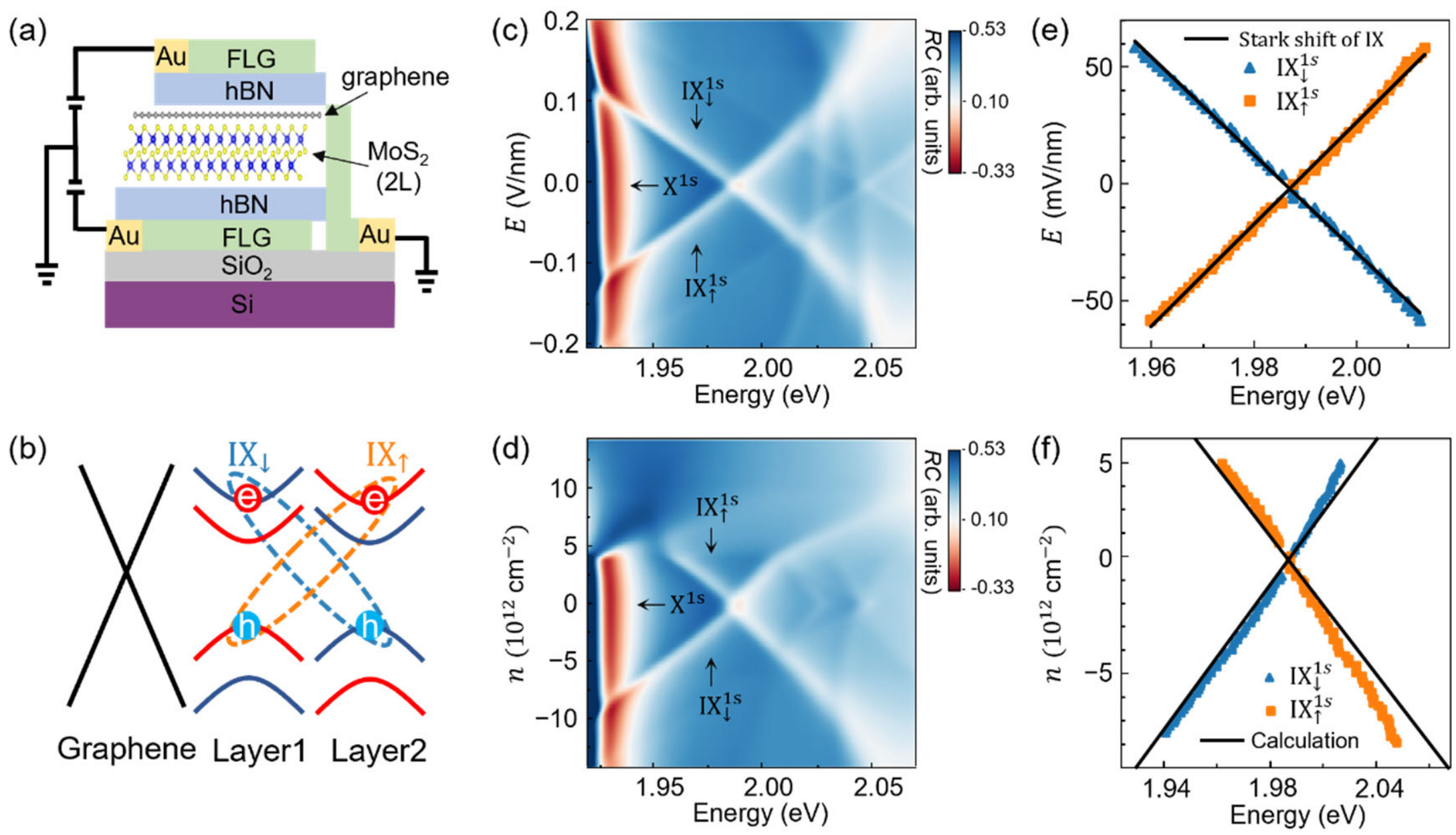

**FIG. 2.** Characterization of interlayer excitons and anomalous Stark shift. (a) Schematic of the dual-gated graphene/bilayer $MoS_2$ device. (b) Band alignment indicating that electrostatically induced carriers reside predominantly in graphene. $IX_{\downarrow}$, $IX_{\uparrow}$ denote interlayer excitons with opposite out-of-plane dipole orientations, in which the electrons are located closer to or farther from the graphene layer, respectively. (c),(d) Reflection-contrast maps as functions of out-of-plane electric field at charge neutrality and carrier density at zero applied gate field, respectively. The 1s interlayer excitons $IX_{\downarrow}^{1s}$ and $IX_{\uparrow}^{1s}$, as well as the 1s intralayer exciton $X^{1s}$, are labeled. (e), (f) Corresponding exciton energies extracted from (c) and (d). The interlayer excitons in (e) exhibit linear Stark shifts consistent with large dipole moments, while the shifts in (f) persist even at zero applied gate field, indicating a charge-transfer-induced internal field not captured by the geometric capacitor model.

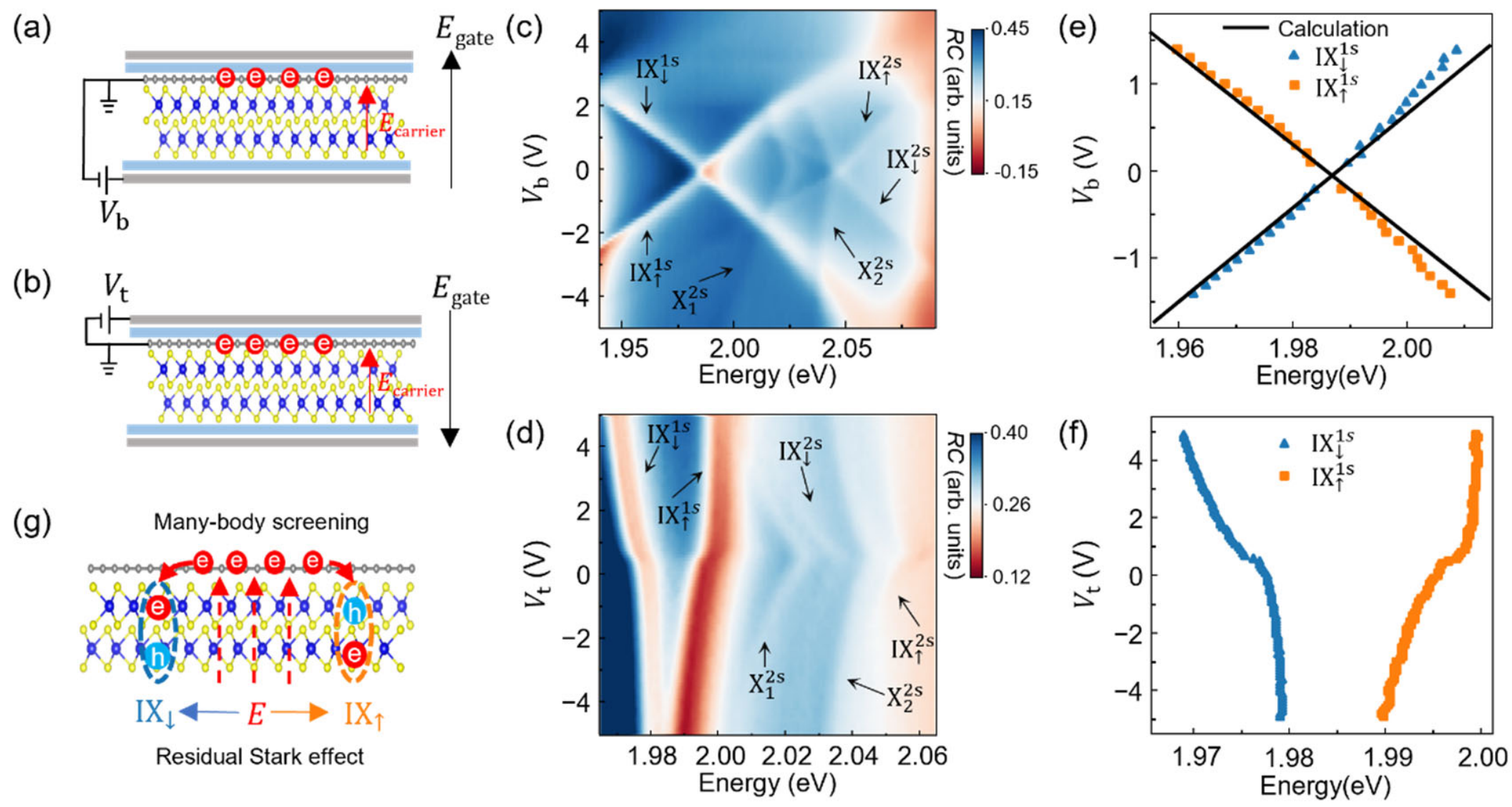

**FIG. 3.** Isolation of nongeometric contributions via single-gate scans. (a), (b) Schematics of the two single-gate configurations. Sweeping the bottom gate $V_b$ produces an external electric field parallel to the carrier-induced internal field, whereas sweeping the top gate $V_t$ yields an antiparallel configuration. (c),(d) Reflection-contrast maps as functions of gate voltage for the bottom-gate scan, with the top-gate voltage fixed at 0 V, and the top-gate scan, with the bottom-gate voltage fixed at −0.6 V, respectively. The 1s and 2s interlayer excitons, as well as the 2s intralayer excitons from the $MoS_2$ layer proximal ($X_1^{2s}$) and distal ($X_2^{2s}$) to graphene, are labeled. (e), (f) Corresponding exciton energies extracted from (c) and (d). The solid lines in (e) show the energy shifts expected from the geometric model. (g) Schematic energy decomposition showing that, after removal of the geometric Stark contribution, the remaining energy shift contains a residual Stark component and many-body screening contributions.

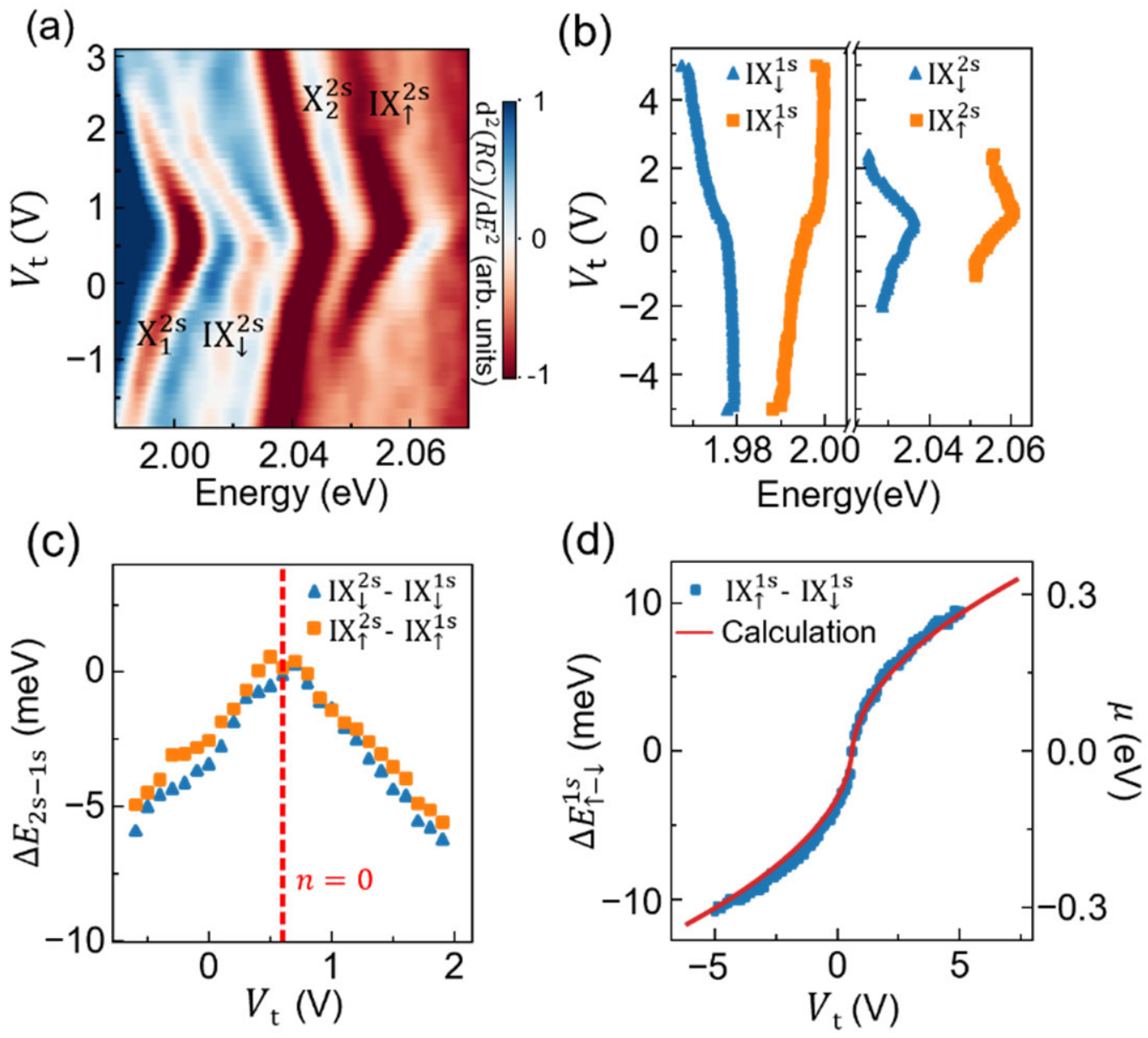

**FIG. 4.** Mechanism identification and chemical potential quantification. (a) Second-derivative reflection-contrast map as a function of top-gate voltage ($V_t$), where the 2s states of both interlayer and intralayer excitons are clearly resolved. (b) Extracted energies of the 1s and 2s interlayer exciton states as a function of $V_t$, showing consistent evolution for both exciton series. (c) Evolution of the $\Delta E_{2s-1s}$ relative to charge neutrality. The nearly identical changes for the two opposite-dipole branches show that the binding-energy renormalization is largely symmetric with respect to dipole orientation and cannot account for the dipole-dependent spectral asymmetry. (d) Change in the energy splitting between the two 1s interlayer exciton branches relative to charge neutrality, as a function of gate voltage. The agreement with the model including graphene quantum capacitance $C_q$ supports the assignment of the observed asymmetry to a residual Stark response. The corresponding splitting provides a direct measure of the graphene chemical potential $\mu$ shown on the right axis.

## Supplemental Material for

# Excitonic Stark Response Induced by Quantum Capacitance in van der Waals Heterostructures

Xiangzhi Hong[1], Yongkang Jian[1], Yang Chen[1], Rukai Cang[1], Yuanlong Guo[1], Takashi Taniguchi[2], Kenji Watanabe[3], Baigeng Wang[1,4,5], Erfu Liu[1,4,5*]

[1]National Laboratory of Solid State Microstructures, School of Physics, Nanjing University, Nanjing 210093, China.

[2]Research Center for Materials Nanoarchitectonics, National Institute for Materials Science, 1-1 Namiki, Tsukuba 305-0044, Japan.

[3]Research Center for Electronic and Optical Materials, National Institute for Materials Science, 1-1 Namiki, Tsukuba 305-0044, Japan.

[4]Collaborative Innovation Center of Advanced Microstructures, Nanjing University, Nanjing 210093, China.

[5]Jiangsu Physical Science Research Center, Nanjing 210093, China.

[*]Corresponding author: erfuliu@nju.edu.cn

## Table of contents

## Note 1. Device fabrication

The graphene/$MoS_2$ heterostructures were fabricated by mechanical exfoliation followed by a polycarbonate (PC)-based dry-transfer technique. An optical microscopy image of the fully assembled Device 1 is shown in Fig. S1(a), with solid outlines marking the boundaries of the constituent flakes.

Monolayer graphene, bilayer $MoS_2$, hexagonal boron nitride (hBN), and few-layer graphite flakes were first exfoliated from bulk crystals onto silicon substrates with a 285-nm $SiO_2$ layer (the $MoS_2$ crystals were obtained from SPI Supplies). The target $MoS_2$ and graphene regions were selected based on their optical contrast and lateral uniformity [Figs. S1(b) and S1(c)]. Atomic force microscopy (AFM) height profiles [Figs. S1(d) and S1(e)] show that the top and bottom hBN layers, which serve as the gate dielectrics, both have a thickness of approximately 13 nm.

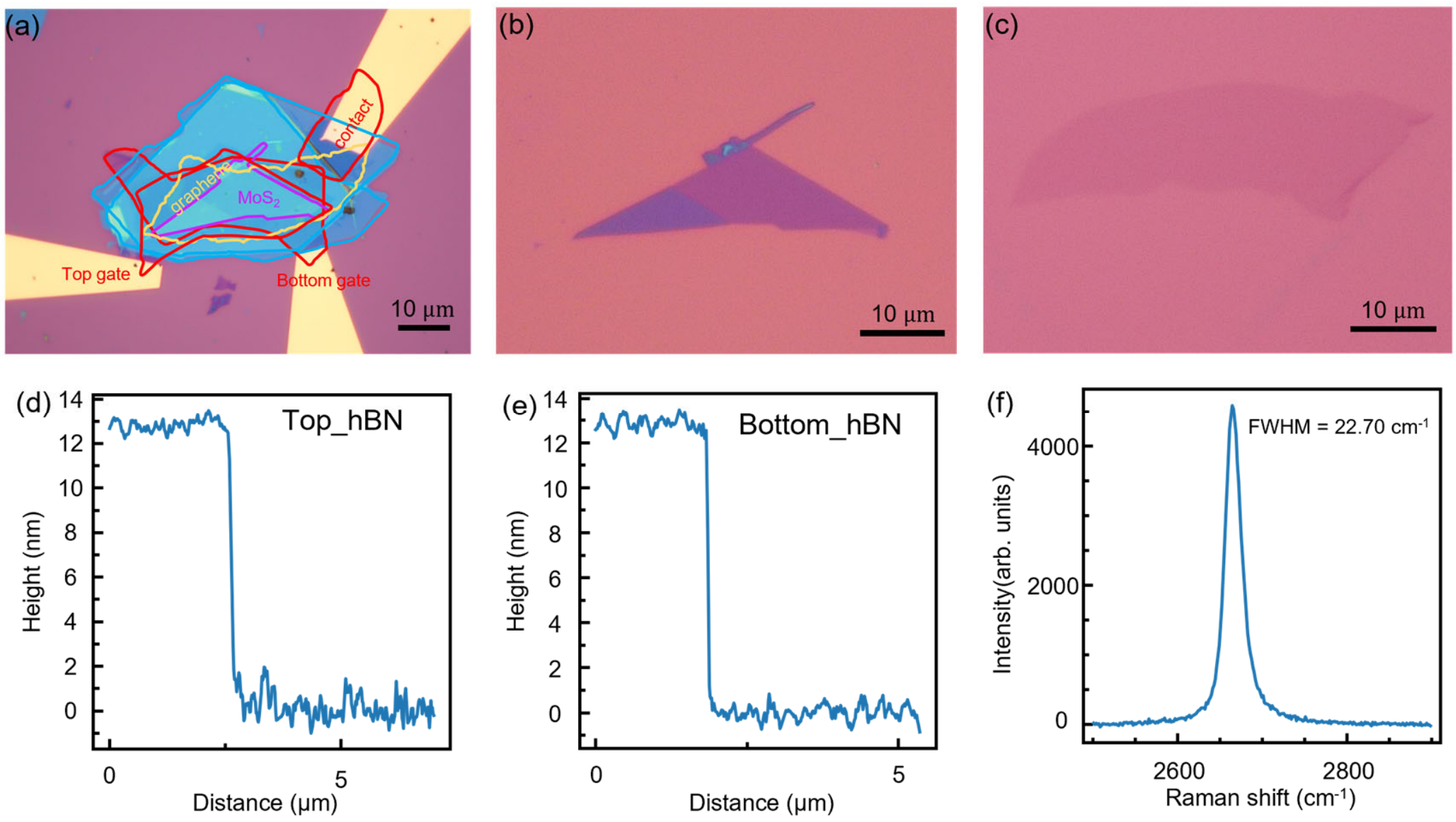

Fig. S1. Fabrication and characterization of the graphene/bilayer $MoS_2$ heterostructure. (a) Optical microscopy image of the fully assembled Device 1. Solid outlines mark the boundaries of the constituent flakes. (b), (c) Optical microscopy images of the isolated $MoS_2$ and monolayer graphene flakes, respectively, before transfer. The $MoS_2$ flake contains both bilayer and thicker multilayer regions. (d), (e) Atomic force microscopy height profiles of the top and bottom hBN flakes, showing a thickness of approximately 13 nm for both dielectric layers. (f) Raman spectrum of the graphene flake measured after device assembly in a region encapsulated only by hBN. The 2D peak is fitted by a single Lorentzian with a full width at half maximum of 22.70 $cm^{-1}$.

The heterostructure was assembled by sequentially picking up the following flakes: few-layer graphite for the bottom gate, hBN, few-layer graphite, graphene, bilayer $MoS_2$, hBN, and few-layer graphite for the top gate. The completed stack was then transferred onto a $SiO_2$/Si substrate. Electrical contacts were defined by electron-beam lithography, followed by thermal evaporation of 55-nm-thick Au onto the few-layer graphite contact regions. The graphene layer number was further examined by Raman spectroscopy in a region of the assembled device where the graphene is encapsulated only by hBN. As shown in Fig. S1(f), the 2D peak is well fitted by a single Lorentzian peak with a full width at half maximum (FWHM) of 22.70 $cm^{-1}$, consistent with monolayer graphene [S1,S2].

**Note 2. Reflection contrast measurements**

The reflection contrast measurements were performed in a vibration-isolated cryostat (Fermi Cryo-300) at a base temperature of 6 K. Broadband white light (Thorlabs SLS302) was coupled into a single-mode fiber and focused onto the sample through a 100× objective (Nikon TU Plan ELWD 100×, NA 0.8), producing a spot size of approximately 1 μm. The reflected light was collected by a spectrometer (Horiba iHR320) equipped with a Syncerity CCD. The reflection contrast (RC) is defined as $RC = (R - R_0) / R_0$, where $R$ is the reflection intensity from the graphene/$MoS_2$ heterostructure, and $R_0$ is the reference reflection intensity measured from a nearby graphene region without $MoS_2$.

**Note 3. Carrier density and electric field from the geometric model**

The carrier density ($n_{\text{geo}}$) and vertical electric field ($E_{\text{geo}}$) are first calculated within a parallel-plate capacitor model. The top-gate capacitance is determined by the top hBN dielectric, $C_{\text{tg}} = \epsilon_0 \epsilon_{\text{hBN}} / d_{\text{top}}$. For the bottom gate, the dielectric stack between graphene and the bottom-gate electrode consists of the bottom hBN layer and bilayer $MoS_2$. The effective bottom-gate capacitance $C_{\text{bg}}$ is therefore given by their series combination:

$$\frac{1}{C_{\text{bg}}} = \frac{1}{C_{\text{hBN}}^{\text{bottom}}} + \frac{1}{C_{\text{MoS}_2}} = \frac{d_{\text{bottom}}}{\epsilon_0 \epsilon_{\text{hBN}}} + \frac{d_{\text{MoS}_2}}{\epsilon_0 \epsilon_{\text{MoS}_2}} \quad \text{(S1)}$$

Here $d_{\text{top}} = 13nm$, $d_{\text{bottom}} = 13nm$ and $d_{\text{MoS}_2} = 1.3nm$ are the thicknesses of the top hBN, bottom hBN, and bilayer $MoS_2$, respectively, while $\epsilon_{\text{hBN}} = 3.1$ and $\epsilon_{\text{MoS}_2} = 6.2$ are their relative dielectric constants.

Within this geometric model, the gate-induced carrier density in graphene is:

$$n = \frac{1}{e}\left(C_{\text{tg}} V_{\text{tg}} + C_{\text{bg}} V_{\text{bg}}\right) \quad \text{(S2)}$$

The geometric electric field $E$ acting on the interlayer excitons inside bilayer $MoS_2$ is obtained from the electric displacement field $D$ divided by the local permittivity $\epsilon_0 \epsilon_{\text{MoS}_2}$:

$$E = \frac{D}{\epsilon_0 \epsilon_{\mathrm{MoS_2}}} = \frac{C_{\mathrm{tg}} V_{\mathrm{tg}} - C_{\mathrm{bg}} V_{\mathrm{bg}}}{2\epsilon_0 \epsilon_{\mathrm{MoS_2}}} \tag{S3}$$

**Note 4. Carrier-induced internal electric field and Stark shift**

Within the geometric model, carriers accumulated in the graphene layer are treated as a two-dimensional charge sheet with areal density $n$. According to Gauss's law, such a charge sheet generates an electric field perpendicular to the interface. The corresponding carrier-induced electric field acting on bilayer $MoS_2$ is written as:

$$E_{\mathrm{carrier}} = \frac{e \cdot n}{2\epsilon_0 \epsilon_{\mathrm{MoS_2}}} \tag{S4}$$

The factor of 1/2 follows from the simplified treatment in which the electric flux generated by the graphene charge sheet is distributed equally on its two sides. This carrier-induced field combines with the gate-induced geometric field and changes the electric field experienced by the interlayer excitons.

For an interlayer exciton with an out-of-plane electric dipole moment $p$, the energy shift varies linearly with the electric field. The expected Stark shift induced by the carrier-generated field is therefore:

$$\Delta E = p \cdot E_{\mathrm{carrier}} = p \cdot \left( \frac{e \cdot n}{2\, \epsilon_0 \epsilon_{\mathrm{MoS_2}}} \right) \tag{S5}$$

Here $p = 0.48e \cdot nm$ is obtained experimentally from the electric-field dependence shown in Fig. 2(e). This expression gives the Stark shift expected when the effect of the transferred carriers is treated entirely within the geometric electrostatic model.

**Note 5. Geometric field cancellation in single-gate scans**

We compare the changes in the gate-induced external field $\Delta E_{\mathrm{gate}}$ and the carrier-induced internal field $\Delta E_{\mathrm{carrier}}$ during a single-gate sweep with gate voltage $V_{\mathrm{g}}$. Within the geometric capacitor model, the change in the gate-induced external field is given by:

$$\Delta E_{\mathrm{gate}} = \pm \frac{C_{\mathrm{g}} \Delta V_{\mathrm{g}}}{2\epsilon_0 \epsilon_{\mathrm{MoS_2}}} \tag{S6}$$

where the $+$ and $-$ signs correspond to the top-gate ($V_{\mathrm{tg}}$) and bottom-gate ($V_{\mathrm{bg}}$) voltages, respectively, with the downward direction defined as the positive electric-field direction.

Simultaneously, the change of the internal field generated by the induced carriers is:

$$\Delta E_{\mathrm{carrier}} = -\frac{e \cdot \Delta n}{2\epsilon_0 \epsilon_{\mathrm{MoS_2}}} = -\frac{e}{2\epsilon_0 \epsilon_{\mathrm{MoS_2}}} \left( \frac{C_{\mathrm{g}} \Delta V_{\mathrm{g}}}{e} \right) = -\frac{C_{\mathrm{g}} \Delta V_{\mathrm{g}}}{2\epsilon_0 \epsilon_{\mathrm{MoS_2}}} \tag{S7}$$

Thus, within the geometric model, $\left|\Delta E_{\text{gate}}\right| = |\Delta E_{\text{carrier}}|$.

In the top-gate configuration, the external geometric field and the carrier-induced internal field are opposite in direction. As a result, the net field change is expected to cancel within the geometric model:

$$\Delta E_{\text{net}} = \Delta E_{\text{gate}} + \Delta E_{\text{carrier}} = 0 \tag{S8}$$

Therefore, within the geometric capacitor model, the top-gate scan configuration leads to a cancellation of the geometric electric-field contribution. As a result, the remaining excitonic energy shifts reflect deviations from the geometric capacitor model, arising from quantum capacitance and many-body effects.

**Note 6. Quantum capacitance-induced Stark splitting**

To account for quantum capacitance, we consider the electrostatics of the dual-gated structure with graphene treated as a conducting sheet with a finite density of states. The electrochemical potential of graphene satisfies $\mu - e \cdot \phi_{\text{graphene}} = 0$, while the bottom gate is treated as an ideal metal with $\phi_{\text{bg}} = V_{\text{bg}}$. The carrier density is then given by:

$$n = \frac{C_{\text{tg}}}{e}\left(V_{\text{tg}} - \frac{\mu}{e}\right) + \frac{C_{\text{bg}}}{e}\left(V_{\text{bg}} - \frac{\mu}{e}\right) \tag{S9}$$

For a fixed bottom-gate voltage $V_{\text{bg}}$, we define $V_{\text{tg}}^0$ as the top-gate voltage corresponding to charge neutrality ($\mu = 0$), which yields $C_{\text{tg}}V_{\text{tg}}^0 + C_{\text{bg}}V_{\text{bg}} = 0$. Using this definition, the density can be rewritten as:

$$n = \frac{C_{\text{tg}}}{e}\left(V_{\text{tg}} - V_{\text{tg}}^0\right) - \frac{\mu}{e^2}\left(C_{\text{tg}} + C_{\text{bg}}\right) \tag{S10}$$

For graphene, the carrier density is related to the chemical potential through the density of states: $n(\mu) = \text{sgn}(\mu)\mu^2/\pi(\hbar v_{\text{F}})^2$. Equating the two expressions for $n$, we obtain:

$$\left(V_{\text{tg}} - V_{\text{tg}}^0\right) = \frac{e}{\pi(\hbar v_{\text{F}})^2 C_{\text{tg}}}\text{sgn}(\mu)\mu^2 + \frac{\left(C_{\text{tg}} + C_{\text{bg}}\right)}{eC_{\text{tg}}}\mu \tag{S11}$$

The electric field across the bilayer $MoS_2$ is determined by the potential difference between graphene and the bottom gate:

$$\Delta\phi = \phi_{\text{graphene}} - \phi_{\text{bg}} = \frac{\mu}{e} - \text{V}_{\text{bg}} = \frac{\mu}{e} + V_{\text{tg}}^0\frac{C_{\text{tg}}}{C_{\text{bg}}} \tag{S12}$$

The corresponding electric field in $MoS_2$ is:

$$E_{\mathrm{MoS2}} = \frac{D}{\epsilon_0 \epsilon_{\mathrm{MoS2}}} = \frac{C_{\mathrm{bg}} \Delta\phi}{\epsilon_0 \epsilon_{\mathrm{MoS_2}}} = \frac{C_{\mathrm{bg}}}{\epsilon_0 \epsilon_{\mathrm{MoS_2}}} \left( \frac{\mu}{e} + V_{\mathrm{tg}}^0 \frac{C_{\mathrm{tg}}}{C_{\mathrm{bg}}} \right) \tag{S13}$$

Referring to the charge-neutrality point ($\mu = 0$), the change in electric field is:

$$\Delta E_{\mathrm{MoS2}} = \frac{C_{\mathrm{bg}}}{e\, \epsilon_0 \epsilon_{\mathrm{MoS_2}}} \mu \tag{S14}$$

For interlayer excitons with dipole moment $p$, the resulting energy splitting is:

$$\Delta E = \frac{2 C_{\mathrm{bg}} p}{e\, \epsilon_0 \epsilon_{\mathrm{MoS_2}}} \mu \tag{S15}$$

Substituting this relation into Eq. (S11), we obtain the dependence of the applied top-gate voltage on the observed Stark splitting:

$$\left( V_{\mathrm{tg}} - V_{\mathrm{tg}}^0 \right) = \frac{e^3 \epsilon_0^2 \epsilon_{\mathrm{MoS_2}}^2}{4\pi (\hbar v_{\mathrm{F}})^2 C_{\mathrm{tg}} C_{\mathrm{bg}}^2 p^2} \mathrm{sgn}(\Delta E) \Delta E^2 + \frac{\left( C_{\mathrm{tg}} + C_{\mathrm{bg}} \right) \epsilon_0 \epsilon_{\mathrm{MoS_2}}}{2 C_{\mathrm{tg}} C_{\mathrm{bg}} p} \Delta E \tag{S16}$$

**Note 7. Experimental calibration**

From the derivation above, the chemical potential $\mu$ is related to the measured Stark splitting $\Delta E$ through a proportionality factor $\Gamma = \frac{2 C_{\mathrm{bg}} p}{e\, \epsilon_0 \epsilon_{\mathrm{MoS_2}}}$.

To determine $\Gamma$ experimentally, we consider the electric-field scan in Fig. 2(c), where the geometric capacitor model remains valid, and charge neutrality is maintained ($C_{tg} V_{tg} + C_{bg} V_{bg} = 0$). In this regime, the total Stark shift arises purely from the geometric electric field. The contributions from the top and bottom gates are:

$$\Delta E_{\mathrm{top}} = -\frac{C_{\mathrm{tg}} p}{\epsilon_0 \epsilon_{\mathrm{MoS_2}}} V_{\mathrm{tg}}, \qquad \Delta E_{\mathrm{bottom}} = \frac{C_{\mathrm{bg}} p}{\epsilon_0 \epsilon_{\mathrm{MoS_2}}} V_{\mathrm{bg}} \tag{S17}$$

Using the neutrality condition, these contributions are equal in magnitude, and the splitting is:

$$\Delta E_{\mathrm{split}} = 2 \frac{C_{\mathrm{bg}} p}{\epsilon_0 \epsilon_{\mathrm{MoS_2}}} V_{\mathrm{bg}} \tag{S18}$$

Compared with the definition of $\Gamma$, we obtain

$$\Gamma = \frac{\Delta E_{\mathrm{split}}}{e V_{\mathrm{bg}}} \tag{S19}$$

This relation enables a direct experimental calibration of the chemical potential from the measured Stark splitting, without relying on dielectric constants or thickness estimates.

**Note 8. Results extracted at other bottom gate voltages**

In the main text, we analyze top-gate voltage sweeps with the bottom gate fixed at

-0.6 V. Here we present the corresponding measurements and calculations for other bottom-gate voltages in Fig. S2. The consistent agreement between experiment and theory demonstrates that the observed behavior is robust and does not depend on the specific choice of bottom-gate voltage.

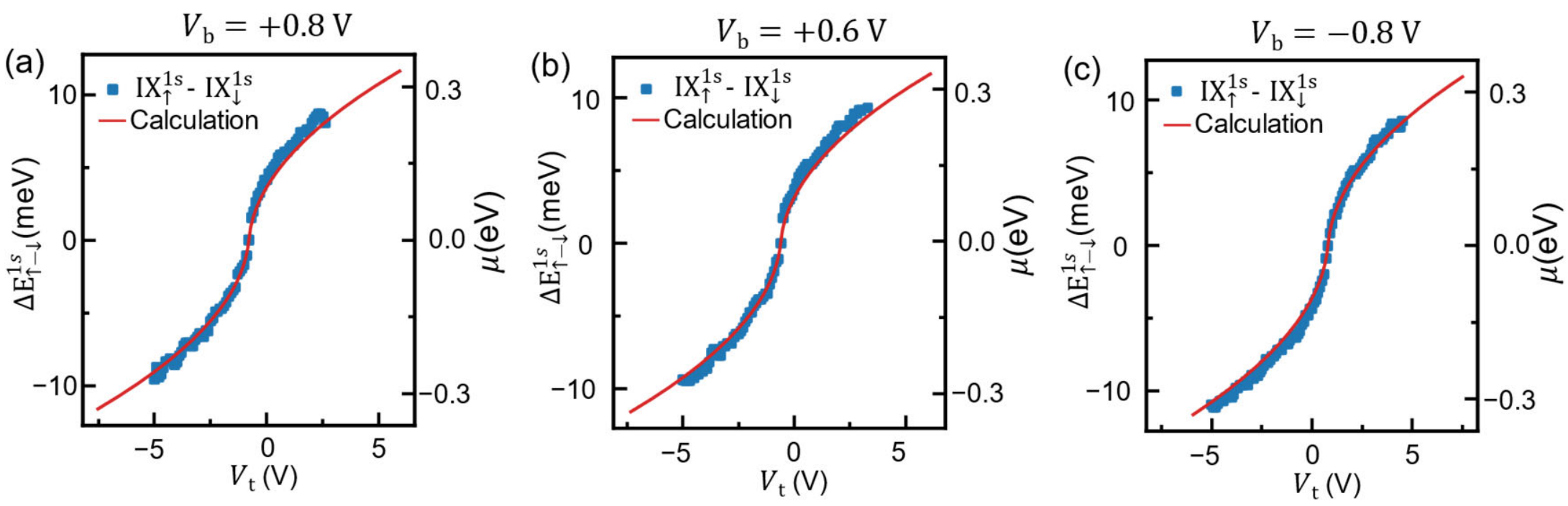

**Fig. S2.** (a–c) Energy splitting between the two interlayer exciton branches (relative to charge neutrality) and the corresponding chemical potential, compared with theoretical calculations, as functions of top-gate voltage for different fixed bottom-gate voltages in Device 1.

**Note 9. Data from Device 2**

Consistent agreement between experiment and theory is observed for Device 2 (Fig. S3), demonstrating the reproducibility of our results across different devices. In Device 2, the top and bottom hBN thicknesses are 21 nm and 23 nm, respectively. Under the same voltage step as in Device 1, the smaller geometric capacitance leads to a reduced Stark splitting, resulting in a lower resolution in extracting the chemical potential under identical measurement conditions.

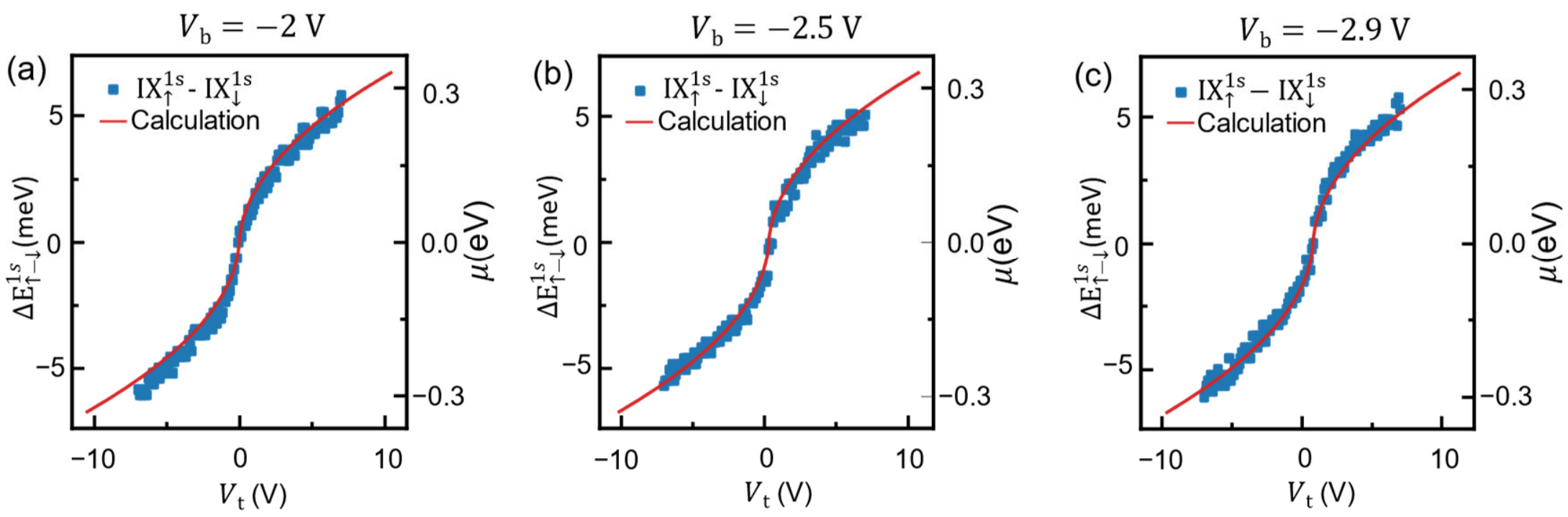

**Fig. S3.** (a–c) Measurements for Device 2: energy splitting between the interlayer exciton branches (relative to charge neutrality) and the extracted chemical potential as functions of top-gate voltage at several fixed bottom-gate voltages, together with theoretical calculation.

**Note 10. Data from Device 3**

To examine how the density of states of graphene affects the residual Stark response,

we measured an additional sample, Device 3, containing separate bilayer-graphene (BLG) and multilayer-graphene (MLG, ~6 nm thick as determined by AFM) regions.

In the BLG region [Figs. S4(a) and S4(b)], the measured interlayer-exciton splitting agrees well with calculations based on the standard four-band tight-binding model of Bernal-stacked bilayer graphene in the low-carrier-density regime, including a displacement-field-induced band gap [S3–S5]. A perpendicular displacement field was converted into an interlayer potential asymmetry $U$, with screening within bilayer graphene treated phenomenologically using an effective dielectric constant $\varepsilon_{\mathrm{eff}} = 3.2$ [S6]. The corresponding band gap was evaluated as:

$$E_{\mathrm{g}} = \frac{|U|\gamma_1}{\sqrt{\gamma_1^2 + U^2}} \tag{S20}$$

Using a Fermi velocity of $v = 8.4 \times 10^5\ \mathrm{m/s}$ and an interlayer hopping parameter of $\gamma_1 = 0.34\ \mathrm{eV}$ [S4], the chemical potential was derived from the resulting gapped four-band dispersion. At low carrier density, the annular Fermi surface associated with the Mexican-hat dispersion was explicitly taken into account. The values of $U$ and $E_{\mathrm{g}}$, evaluated at the charge-neutrality point, were held fixed throughout the calculation. Trigonal warping, interaction corrections, and self-consistent carrier-density-dependent screening were neglected.

In the MLG region [Fig. S4(c)], the splitting shows only a weak dependence on the top-gate voltage within our experimental resolution. This behavior is consistent with the larger density of states of multilayer graphene, for which the quantum capacitance is much larger than the relevant geometric capacitance over the measured voltage range. Therefore, its chemical potential exhibits negligible shift during the top-gate sweep, meaning the internal electric field acting on the interlayer excitons remains virtually unchanged.

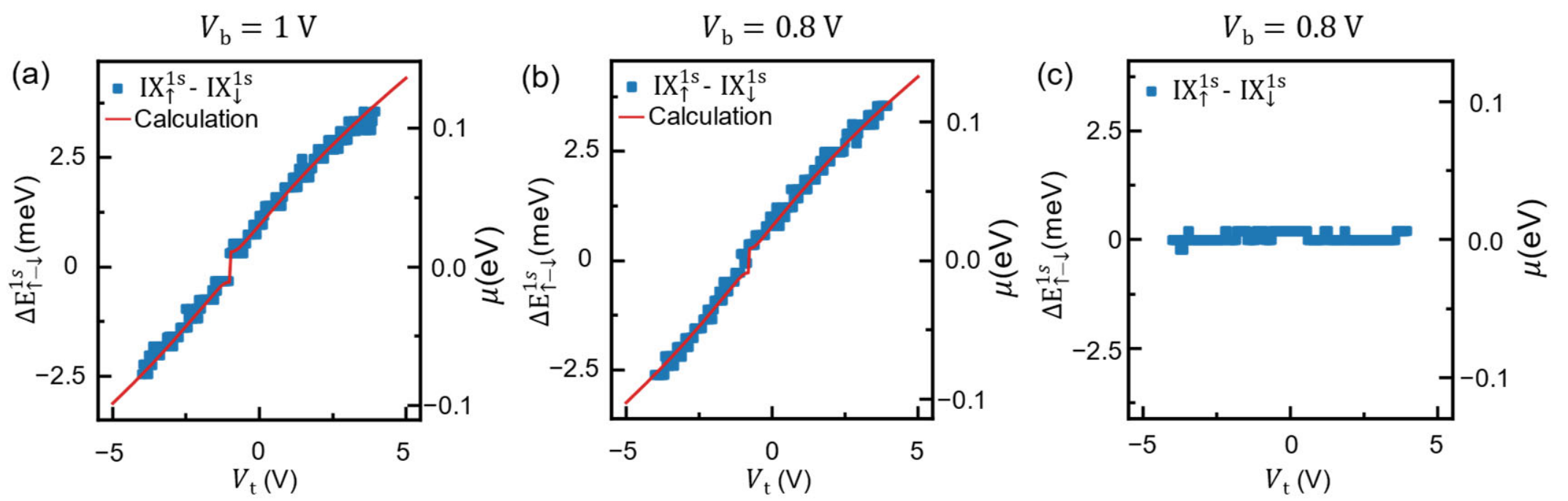

**Fig. S4.** Measurements from Device 3. Energy splitting between the two interlayer exciton branches and the extracted chemical potential as functions of top-gate voltage at fixed bottom-gate voltages. (a), (b) Experimental data (blue squares) and theoretical calculations (red lines) for the bilayer graphene region at $V_{\mathrm{b}} = 1$ V and $V_{\mathrm{b}} = 0.8$ V, respectively. (c) Experimental data for the multilayer graphene region at $V_{\mathrm{b}} = 0.8$ V. The splitting shows only a weak dependence on top-gate voltage within the experimental resolution, consistent with the larger density of states of multilayer graphene.

**Note 11. Many-body screening from the dipole-symmetric component**

In Fig. 4(d) of the main text, the local chemical potential of graphene is obtained from the energy splitting between the two interlayer exciton branches, which represents the dipole-dependent component. In contrast, the many-body screening effect is captured by the shift in the average energy of the interlayer excitons (the dipole-symmetric component), as shown in Fig. S5.

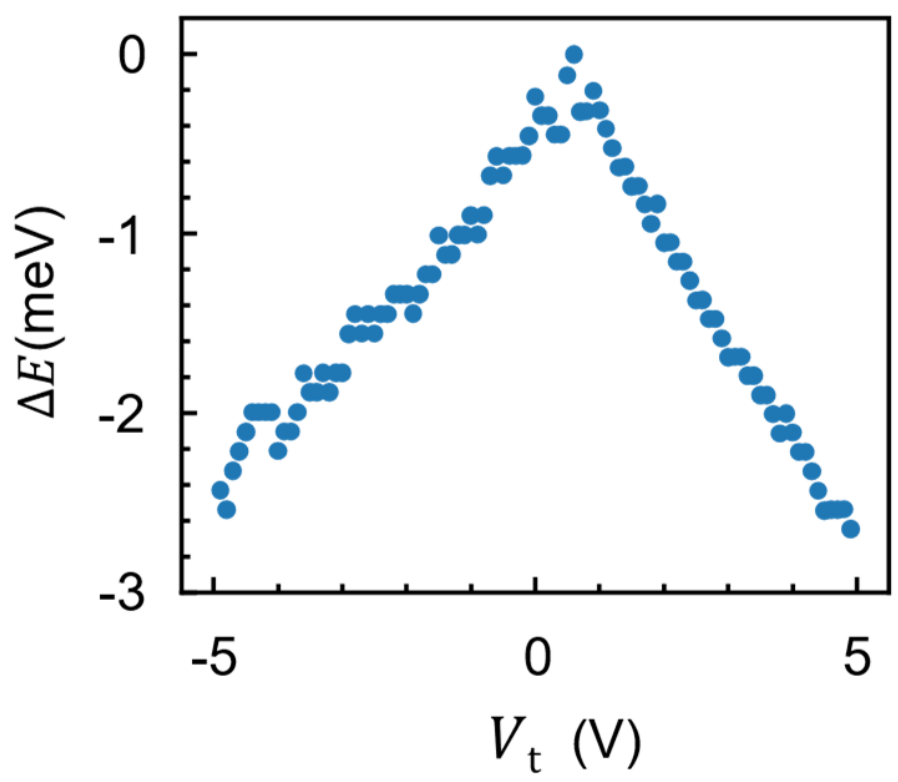

**Fig. S5.** Energy shift of the interlayer excitons associated with many-body screening as a function of top-gate voltage.

**Supplementary References:**